\documentclass[pre,amsmath,amssymb,twocolumn]{revtex4}

\usepackage{amsmath}    
\usepackage{amsfonts}
\usepackage{amssymb}
\usepackage{graphicx}   
\usepackage{dcolumn}
\usepackage{bm}

\begin{document}

\title{Robustness of the filamentation instability as shock mediator in arbitrarily oriented magnetic field}

\author{A. Bret}
 \email{antoineclaude.bret@uclm.es}

\author{E. Perez Alvaro}
 \email{erica.perez@uclm.es}

\affiliation{ETSI Industriales, Universidad de Castilla-La Mancha, 13071 Ciudad Real, Spain}
\affiliation{Instituto de Investigaciones Energéticas y Aplicaciones Industriales, Campus Universitario de Ciudad Real, 13071 Ciudad Real, Spain}

\date{\today }

\begin{abstract}
The filamentation instability (sometimes also referred to as ``Weibel'') is a key process in many astrophysical scenario. In the Fireball model for Gamma Ray Bursts, this instability is believed to mediate collisionless shock formation from the collision of two plasma shells. It has been known for long that a flow aligned magnetic field can completely cancel this instability. We show here that in the general case where there is an angle between the field and the flow, the filamentation instability can never be stabilized, regardless of the field strength. The presented model analyzes the stability of two symmetric counter-streaming cold electron/proton plasma shells. Relativistic effects are accounted for, and various exact analytical results are derived. This result guarantees the occurrence of the instability in realistic settings fulfilling the cold approximation.
\end{abstract}


\maketitle

\section{Introduction}
Gamma Ray Bursts (GRB's) and High Energy Cosmic Rays (HECR's) are two of the most intriguing enigmas in contemporary astrophysics \citep{Bahcall97}. A particulary promising scenario explaining both problems consists in the so-called ``Fireball model'' where particles are accelerated through Fermi-type shock acceleration \cite{Achterberg2001} in a relativistic shock generated by collisions between various shells of a relativistic ejecta \citep{Sari1997,Piran2004}. Highly energetic particles escaping the shock environment ends up as HECR's, while those still undergoing Fermi cycles near the shock front radiate in the $\gamma$ range, explaining the GRB emission \citep{MedevedevJitter2000,KirkReville}.

This scenario has been validated in recent years through a series of pioneering Particle-In-Cell (PIC) Simulations where the very shock formation, followed by Fermi like particle acceleration, has been directly observed \citep{SilvaApJ,Spitkovsky2008a}. A key part of the proposed mechanism is the very internal shock formation from the collision of similar density plasma shells, which is believed to be mediated by the filamentation instability occurring when collisionless plasma shells are crossing each others \citep{Medvedev1999}.

The filamentation instability studied here, namely unstable modes propagating perpendicularly to the flow of two counter-streaming species, is frequently labeled ``Weibel'' in literature \cite{Cary1981,califano1,Medvedev1999,Silva2002}, although differences exist between these two \cite{BretPoPReview}. When considering a flow-aligned magnetic field, this instability has been alternatively called ``filamentation'' \cite{Molvig} or ``Weibel'' \cite{Cary1981}. Some connections could equally be drawn with the so-called mirror mode instability \cite{Gedalin1993,Gedalin2001}. The mirror-mode instability setting has common points with the magnetized version of the Weibel instability \cite{LazarPoP2009}. In both cases, the unstable system consists in an anisotropic plasma immersed in an external magnetic field. In our case, the instability arises from the relative drift of two \emph{isotropic} species in their own rest frame. Weibel and filamentation can intertwine when considering relatively drifting \emph{anisotropic} species \cite{lazarPoP2006,lazarPoP2008}.

In view of its central role, much attention has been devoted in evaluating the robustness of the filamentation instability. It has been found for example that temperature effects can significantly reduce its growth rate, and even cancel it \citep{Cary1981,Silva2002}. In the same perspective, it has been determined that a flow aligned magnetic field can suppress the instability \citep{Godfrey1975,StockemApJ,StockemPPCF}, while baryon loading of the plasma shells can increase the growth rate even for large temperatures \citep{FioreMNRAS}.

\begin{figure*}
\begin{center}
 \includegraphics[width=0.25\textwidth]{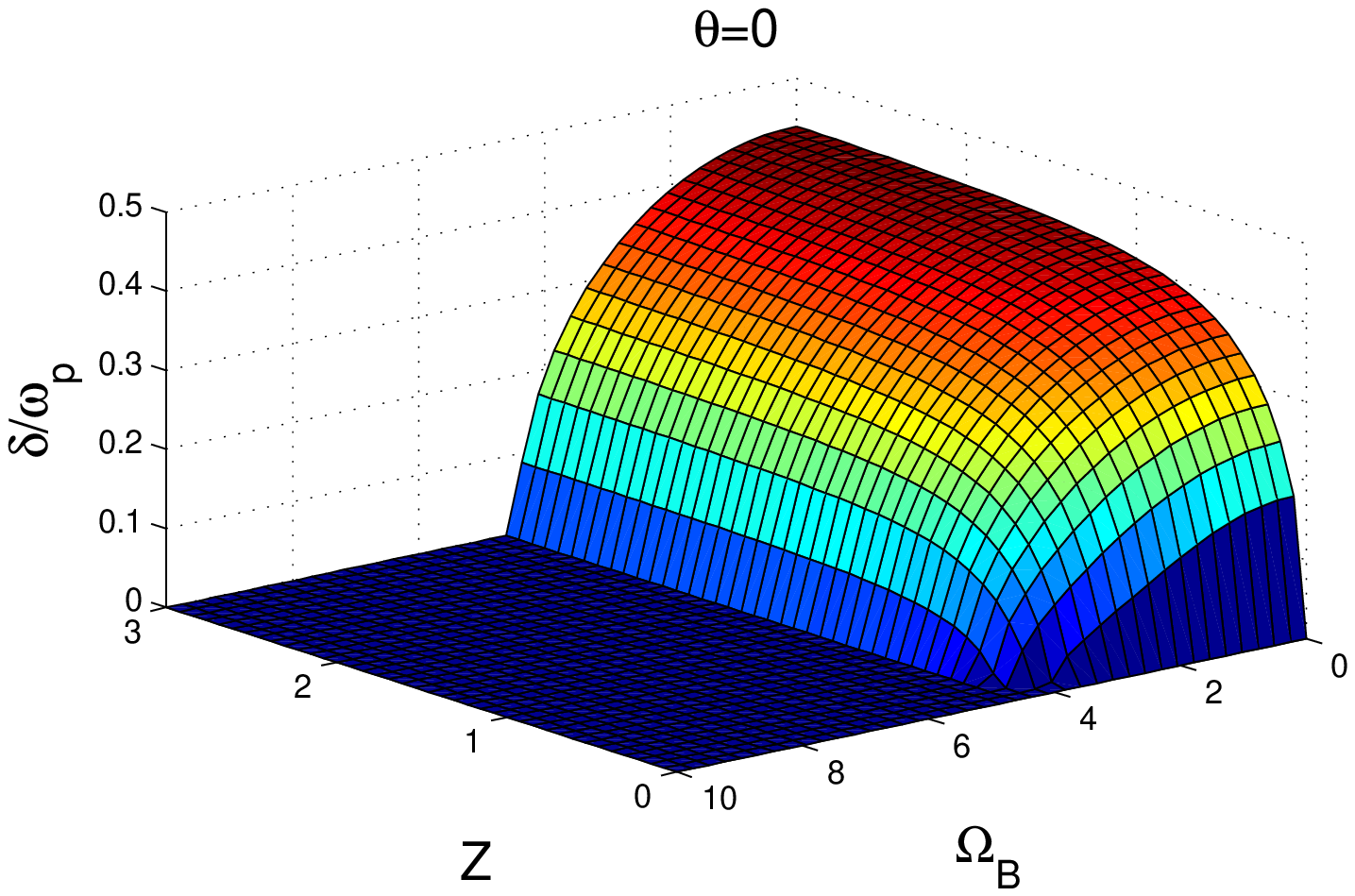}~~~ \includegraphics[width=0.25\textwidth]{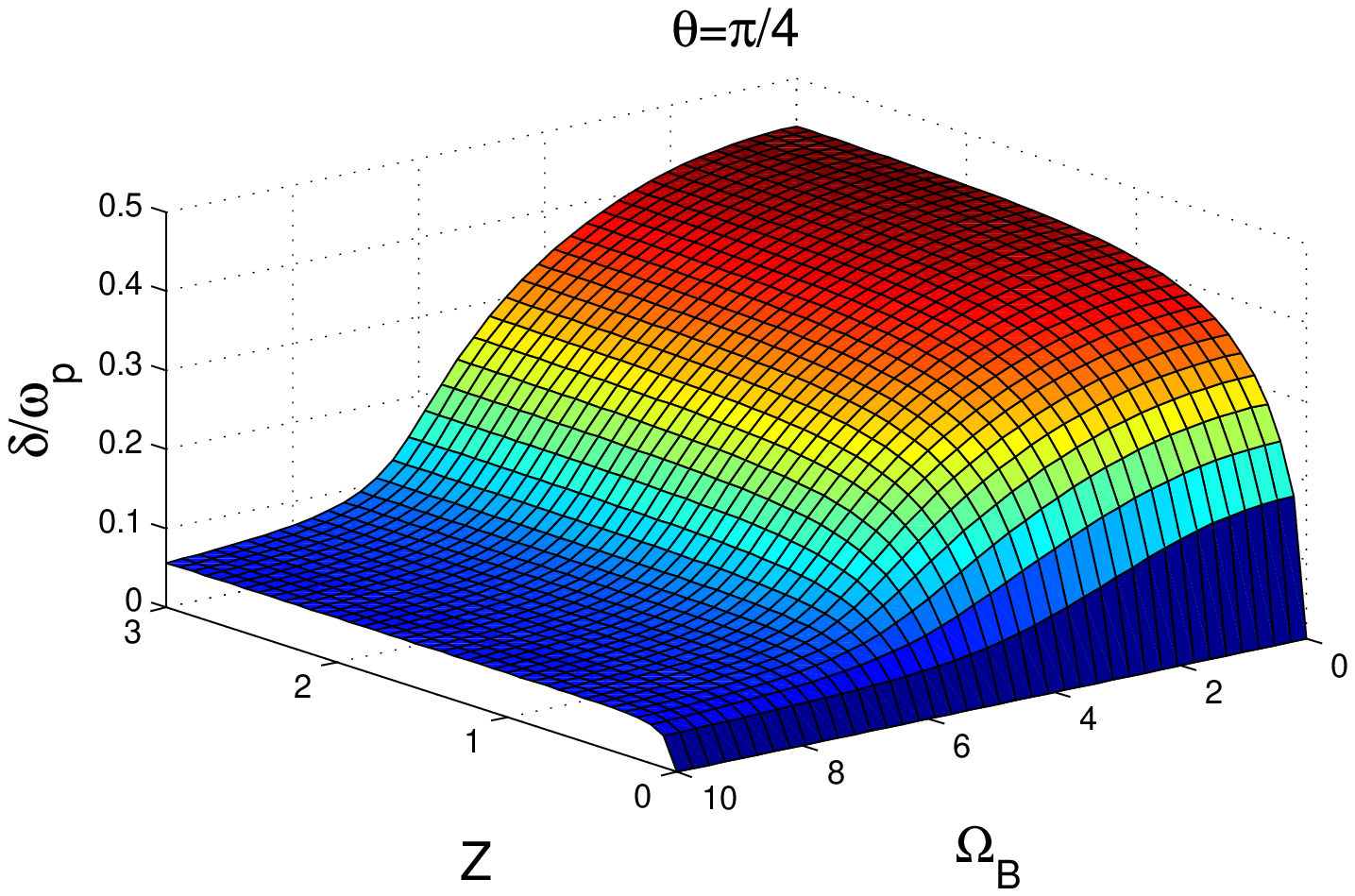}~~~ \includegraphics[width=0.25\textwidth]{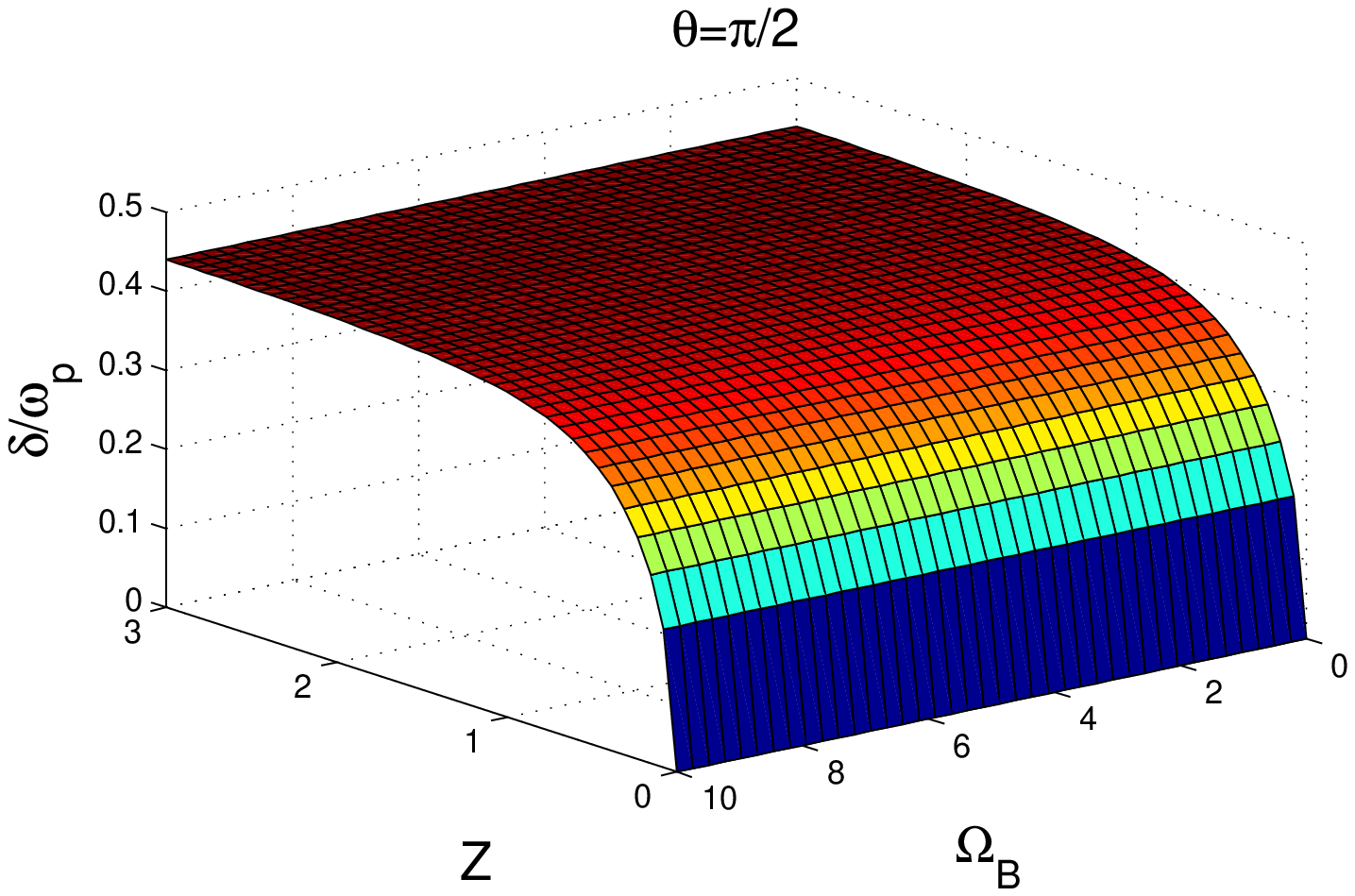}
\end{center}
\caption{(Color online) Growth rate $\delta$ ($\omega_p$ units) in terms of $Z=kv_b/\omega_p$ and $\Omega_B=\omega_b/\omega_p$ for various $\theta$. The Lorentz factor is $\gamma=10$.}
\label{fig:1}
\end{figure*}

The filamentation instability is far from being the only unstable mode triggered by counter-streaming plasmas \citep{BretPRL2005}. Nevertheless, it is the only kind of mode producing  electromagnetic turbulence because the rest of the unstable spectrum of such systems is mainly electrostatic \citep{fainberg,BretPRE2010}. The occurrence of the instability is thus desirable if one wishes to witness the growth of magnetic fields, even if it does not dominate the unstable spectrum.

Although the angle $\theta$ between an hypothetical external magnetic field $\mathbf{B}_0$ and the initial velocity of the colliding shells, is likely to be non-zero in a realistic situation, numerical or analytical works usually consider $\theta=0$ besides a few exceptions \citep{BretPoPOblique,SironiApj}. The goal of this Letter is precisely to explore analytically the general case of a non flow-aligned magnetic field. In order to simplify the analytical treatment, we focus on the role of the filamentation instability in the internal shock formation phase by considering the collision of two identical relativistic cold electron/proton plasma shells. A density ratio different from unity would be worth exploring, as important consequences on the growth rate have been reported \cite{Godfrey1975,Michno2010}. Leaving the asymmetric case for later works, the present theory can still form a support for many computational works where the encounter of identical plasma shells is considered \cite{Kazimura,SilvaApJ,Jaroschek2005,Spitkovsky2008,Spitkovsky2008a}. The main result, that we think is important for the setting considered, is that the magnetic field can cancel the instability only for $\theta=0$. For $\theta\neq 0$, the instability growth rate do decline with $B_0$, but reaches a finite limit for $B_0\rightarrow\infty$. The occurrence of the instability is thus assured, at least in the cold limit.

\section{Formalism and Dispersion Equation}
We consider the system formed by two infinite identical counter-streaming electron/proton plasma shells of density $n$, velocity $\pm \mathbf{v}_b$ and Lorentz factor $\gamma=(1-v_b^2/c^2)^{-1/2}$. We choose to align the flow with the $z$ axis and set the magnetic field $\mathbf{B}_0$ in the $(x,z)$ plane with $(\widehat{\mathbf{e}_z,\mathbf{B}_0})=\theta$. We neglect in each shell the protons inertia, given their much higher mass and their Lorentz factor equal to that of the electrons. We work from the cold fluid and momentum equations for each electron species $j=1,2$,
\begin{equation}\label{eq:consev}
    \frac{\partial n_j}{\partial t}+\nabla\cdot(n_j\mathbf{v}_j)=0,
\end{equation}
\begin{equation}\label{eq:euler}
    \frac{\partial \mathbf{p}_j}{\partial t}+(\mathbf{v}_j\cdot\nabla)\mathbf{p}_j=q\left[\mathbf{E}+\frac{\mathbf{v}_j\times(\mathbf{B}+\mathbf{B}_0)}{c}\right],
\end{equation}
where  $\mathbf{p}_j=\gamma_jm\mathbf{v}_j$ is the momentum and electric charge of specie $j$, and $q,m$ the electron charge and mass respectively. Linearizing these equations and closing the system with Maxwell's equations allows to derive the dielectric tensor under the form,
\begin{equation}\label{eq:tensor}
  \mathcal{T}=\left(%
\begin{array}{ccc}
  T_{xx}    &  T_{xy}    &   T_{xz}  \\
  T_{xy}^*  &  T_{yy}    &   T_{yz}  \\
  T_{xz}^*  &  T_{yz}^*  &   T_{zz}  \\
\end{array}%
\right),
\end{equation}
where $z^*$ is the complex conjugate of $z$, while the tensor elements are reported in Appendix \ref{app:tensor} in terms of the following dimensionless variables,
\begin{equation}\label{eq:variables}
  x = \frac{\omega}{\omega_p}, ~~Z=\frac{k v_b}{\omega_p},~~
  \beta=\frac{v_b}{c},~~\Omega_B = \frac{\omega_b}{\omega_p}.
\end{equation}
Here, $\omega_p^2=4\pi n q^2/m$ is the electronic plasma frequency of a single shell, and $\omega_b=|q|B_0/mc$ the non-relativistic electronic cyclotron frequency. The dispersion equation to be solved is obtained equalling to 0 the determinant of the tensor $\mathcal{T}$. Figure \ref{fig:1} displays the numerical evaluation of the growth rate in terms of the reduced wave vector $Z$ and the magnetic field parameter $\Omega_B$ for various obliquity angles $\theta$. We recover previous known results for the flow-aligned case, as the growth rate saturates to a finite value  for $Z\rightarrow\infty$, while the entire spectrum is stabilized beyond $\Omega_B=\beta\sqrt{2\gamma}$ \citep{BretPoPMagne}. At $\theta=\pi/4$, the large $Z$ saturation is maintained, but the magnetic field effect is qualitatively different: although the growth-rate drops below a critical magnetization, it does not drop to zero but seems to saturate to a finite value. Finally, the case $\theta=\pi/2$ hardly displays any effect of the magnetic field. We will now see that it is possible to analytically access much of the quantities involved here.

\section{Analytical expression for the maximum growth rate}
\begin{figure}
\begin{center}
 \includegraphics[width=0.45\textwidth]{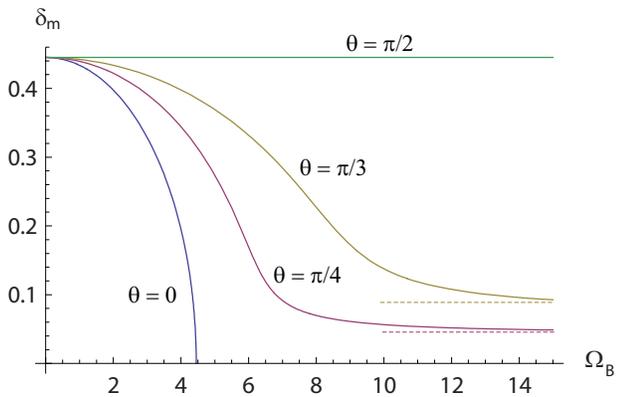}
\end{center}
\caption{(Color Online) Numerical evaluation of the maximum growth rate $\delta_m$ of the filamentation instability in terms of the magnetization parameter $\Omega_B$ for various field orientations and $\gamma=10$. The dotted lines indicate the values given by Eq. (\ref{eq:taux_large}). The green dotted line is coincident with the plain one for $\theta=\pi/2$, and Eq. (\ref{eq:taux_large}) gives $\delta_m^\infty=0$ for $\theta=0$.}
\label{fig:3}
\end{figure}

As can be checked  on Figure \ref{fig:1},  for all values of $\Omega_B$ the growth rate quickly reaches a constant value for large $Z$, and this constant is the maximum growth rate for all $Z$'s. Note that this $Z$-saturation of the growth rate is \emph{observed}, not \emph{demonstrated}. Such behavior of the filamentation instability for cold systems, magnetized or not, has already been reported  \cite{Godfrey1975,califano1}. We thus focus from now on the value of the growth rate for $Z=\infty$. The dispersion function, though lengthy, is eventually a polynomial of degree $d$ in $Z$ which can be written under the form,
\begin{equation}\label{eq:P}
P(Z)=\sum_{m=0}^d a_mZ^m = Z^d\left(a_d+\sum_{m=0}^{d-1} a_mZ^{m-d}\right).
\end{equation}
The right-hand-side of this equation shows that for $Z\rightarrow\infty$, $P(Z)=0$ is equivalent to $a_d=0$, since the $a_i$'s don't depend on $Z$ and $m-d\leq-1$ for $0\leq m\leq d-1$. Surprisingly, the coefficient $a_d$ can be factorized exactly as $a_d=Q_1Q_2Q_3$ with,
\begin{eqnarray}\label{eq:disper_factors}
  Q_1 &=& (x^2\gamma -2) (2 x^2 \gamma^4-2\Omega_B^2\sin^2\theta)-2 x^2 \gamma^3 \Omega _B^2\cos^2 \theta, \nonumber\\
  Q_2 &=& (x^2\gamma+2 \beta^2) (2 x^2 \gamma^4 -2\Omega_B^2\sin^2\theta)-2 x^2\gamma ^3 \Omega _B^2\cos^2\theta,\nonumber\\
  Q_3 &=& x^2 \gamma ^4-(\gamma ^2 \cos ^2\theta+\sin^2\theta) \Omega _B^2.
\end{eqnarray}
It is obvious that $Q_3=0$  only yields stable modes. We prove in Appendix \ref{app:proof} that $Q_1$  only yields stable modes, while $Q_2$ always yields one purely growing mode with $\mathcal{R}\omega=0$, unless $\theta=0$. It has been known for long that the filamentation instability can be canceled by a flow aligned magnetic field. We here find that this result holds only for such field orientation. The dispersion equation $Q_2=0$ can be solved exactly, yielding the $Z$-asymptotic, and maximum, growth rate $\delta_m$ as,
\begin{eqnarray}\label{eq:taux}
 \delta_m^2&=&\frac{b+\sqrt{\Delta}}{2a},~~\Delta=b^2-4ac,~~a = 2\gamma^5, \nonumber\\
 b &=& \gamma\left(4 \beta ^2 \gamma^3-\Omega_B^2+\Omega_B^2 (\cos2\theta-2\gamma^2\cos^2\theta)\right),\nonumber\\
 c &=& -4 \beta ^2 \Omega_B^2 \sin^2\theta.\nonumber\\
\end{eqnarray}
For $\Omega_B=0$, the known result is retrieved with $\delta_m=\beta\sqrt{2/\gamma}$. Setting $\theta=\pi/2$ in the equations above yields  $\delta_m=\beta\sqrt{2/\gamma}$ regardless of $\Omega_B$, showing (as hinted by Fig. \ref{fig:1}) that the maximum growth rate in the case of a perpendicular magnetic field no longer depends on the field amplitude. Figure \ref{fig:3} shows the variations of $\delta_m$ in terms of $\Omega_B$ for various field orientations. At $\theta=0$, the suppression of the instability is recovered for $\Omega_B>\beta\sqrt{2\gamma}$ \citep{BretPoPMagne}. The absence of any field effect for $\theta=\pi/2$ is confirmed, while the curves at intermediate orientation show the system undergoes a transition around a critical $\Omega_B^*$. For $\Omega_B\ll \Omega_B^*$, the system behaves similarly to the $\theta=0$ case. But for $\Omega_B\gg\Omega_B^*$, a steady growth rate is reached, independent of the magnetization.

The following expansion is readily derived for low magnetization,
\begin{equation}\label{eq:expan_small}
  \delta_m \sim \beta\sqrt{\frac{2}{\gamma}}\left(1-\frac{\cos^2\theta}{4\beta^2\gamma}\Omega_B^2\right),
\end{equation}
showing the critical magnetic parameter $\Omega_B^*$ is,
\begin{equation}\label{eq:critique}
\Omega_B^*=\frac{2\beta\sqrt{\gamma}}{\cos\theta}.
\end{equation}
For large field $\Omega_B\gg \Omega_B^*$, the growth rate approaches the limit,
\begin{equation}\label{eq:taux_large}
\delta_m^\infty = \beta\sqrt{\frac{2}{\gamma}}\frac{1}{\sqrt{1+\gamma^2 \cot^2\theta}},
\end{equation}
which is simply the maximum growth rate for $\theta=0$ times a correction factor accounting for the geometry. In relativistic conditions, the critical field parameter defined by Eq. (\ref{eq:critique}) is always larger than 1.

\section{Discussion and Conclusion}
We thus find that within the limits of the cold regime (see below), the occurrence of the instability is ensured as long as the field is not strictly flow aligned. Shock formation mediated by the filamentation instability seems therefore quite robust. Besides the cold approximation, we also neglected the proton inertia. Accounting for the later is unlikely to result in a lesser unstable system, as it has so far been found that Baryon inertia rather sustains the instability, even in the kinetic regime \citep{FioreMNRAS}.

The main limitation of this work lies therefore in the cold approximation.  To be valid, the transverse velocity spread with respect to the flow $\Delta v_\perp$ must fulfills \citep{fainberg}
\begin{equation}\label{eq:cold}
 k^{-1} \gg \Delta v_\perp \delta_m^{-1},
\end{equation}
where $k$ is the typical filamentation wavelength. This inequality simply states that particles can be considered as monokinetic during one growth period, from the standpoint of the plane wave propagating normally to the flow \citep{BretPoPReview}. Taking $k\sim c/\omega_p$ and considering Maxwell-J\"{u}ttner distribution functions for which $\Delta v_\perp/c\sim (2k_BT/\gamma m c^2)^{1/2}$ \citep{BretPRE2010} yields,
\begin{equation}\label{eq:cold1}
 k_BT \ll \frac{\gamma}{2}\delta_m^2 mc^2.
\end{equation}
For weak magnetization with $\Omega_B\ll \Omega_B^*$, $\delta_m\sim\sqrt{2/\gamma}$ simply gives $k_BT \ll mc^2$, implying non-relativistic temperatures. In the strong field limit $\Omega_B\gg \Omega_B^*$, the condition $k_BT \ll mc^2/(1+\gamma^2 \cot^2\theta)$ can be more stringent.

The Maxwell-J\"{u}ttner distribution does not allow for a separate definition of a transverse and  a parallel temperature \cite{LazarOPPJ2010}. Considering such distribution function like the waterbag, we have $\Delta v_\perp\sim \Delta P_\perp/m\gamma$ \citep{BretEPL}, yielding a limitation on the \emph{transverse} momentum spread only,
\begin{equation}\label{eq:water}
 \Delta P_\perp \ll \delta_m \gamma mc
 =
\left\{ \begin{array}{ll}
\sqrt{2\gamma}mc, & \Omega_B\ll \Omega_B^*,\\
\sqrt{2\gamma}mc/{\sqrt{1+\gamma^2 \cot^2\theta}}, & \Omega_B\gg \Omega_B^*.
\end{array} \right.
\end{equation}
Note that for any function of the form $f(\mathbf{v})=f_\perp(v_\perp,T_\perp)f_\parallel(v_\parallel,T_\parallel)$, the transverse spread $\int v_\perp f d^3v$ only depends on $T_\perp$. As it appears, a direct application of the present results depends of the kind of distribution function considered. A fully relativistic kinetic theory would then be needed to assess the robustness of the instability for oblique propagation when the inequalities above are not fulfilled.

\acknowledgments
This work has been  achieved under projects ENE2009-09276 of
the Spanish Ministerio de Educaci\'{o}n y Ciencia and
PEII11-0056-1890 of the Consejer\'{i}a de Educaci\'{o}n y Ciencia de
la Junta de Comunidades de Castilla-La Mancha.

\begin{figure}[t]
\begin{center}
 \includegraphics[width=0.45\textwidth]{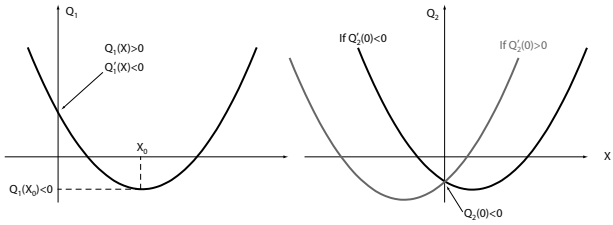}
\end{center}
\caption{Qualitative sketch of the polynomial curves $Q_1(X)$, left, and $Q_2(X)$, right.}
\label{fig:4}
\end{figure}

\appendix

\section{Tensor elements}\label{app:tensor}
We report here the elements of the dielectric tensor defined in Eq. (\ref{eq:tensor}),
\begin{widetext}
\begin{eqnarray}
  T_{xx} &=& \frac{\left(x^2 \gamma -2\right) \left(2 x^2 \gamma ^4+\Omega _B^2\cos2\theta-\Omega _B^2\right)-2 x^2 \gamma ^3 \Omega _B^2\cos ^2\theta}{2 x^2
   \gamma ^5-2 \gamma  \left(\gamma ^2 \cos ^2\theta+\sin ^2\theta\right) \Omega _B^2}, \\
  T_{yy} &=& x^2 \left(1-\frac{2 \gamma ^3}{x^2 \gamma ^4-\left(\gamma ^2 \cos ^2\theta+\sin ^2\theta\right) \Omega _B^2}\right)-\frac{Z^2}{\beta ^2},\\
  T_{zz} &=& x^2-\frac{Z^2}{\beta ^2}+\frac{-2 x^2 \left(x^2+Z^2 \gamma ^2\right) \gamma ^2+\left(x^2+Z^2\right) \Omega _B^2+\left(x^2-Z^2\right) \Omega
   _B^2\cos2\theta}{\gamma  \left(x^2 \gamma ^4-\left(\gamma ^2 \cos ^2\theta+\sin ^2\theta\right) \Omega _B^2\right) x^2},\\
  T_{xy} &=& -\frac{2 i x \gamma ^2 \Omega _B\cos \theta}{x^2 \gamma ^4-\left(\gamma ^2 \cos ^2\theta+\sin ^2\theta\right) \Omega _B^2}, \\
  T_{xz} &=&  \frac{\Omega _B^2\sin2\theta}{x^2 \gamma ^5-\gamma  \left(\gamma ^2 \cos ^2\theta+\sin ^2\theta\right) \Omega _B^2},\\
  T_{yz} &=& -\frac{2 i x \Omega _B\sin \theta}{x^2 \gamma ^4-\left(\gamma ^2 \cos ^2\theta+\sin ^2\theta\right) \Omega _B^2},~~\mathrm{with}~~i^2=-1.
\end{eqnarray}
\end{widetext}

\section{Proof that only $Q_2=0$ yields unstable modes in Eqs. (6)}\label{app:proof}
It is appropriate to set $X\equiv x^2$ in $Q_1$ and $Q_2$, and to study the resulting second order equation in $X$. For both polynomials, the coefficient of $X^2$ is $2\gamma^5>0$. They thus both represent parabolas tending to $+\infty$ for $X\rightarrow\pm\infty$.

For the polynomial $Q_1(X)$, we have $Q_1'(0)=-\gamma\left(4\gamma^3+(1+\gamma^2)\Omega_B^2+(\gamma^2-1)\Omega_B^2\cos2\theta\right)<0$, and $Q_1(0)=4 \Omega_B^2 \sin^2\theta>0$. It is also straightforward to show that the extremum $Q_1(X_0)$, where  $Q_1'(X_0)=0$, is always strictly negative. The curve representing $Q_1$ is thus necessarily qualitatively similar to the one pictured on Fig. \ref{fig:4}. The equation $Q_1(X)=0$ has therefore two strictly positive real roots, so that $Q_1(x)=0$ has only purely real roots.

Regarding $Q_2(X)$, we have $Q_2(0)=-4 \beta^2 \Omega_B^2 \sin^2\theta<0$. Depending on the sign of $Q_2'(0)$, the curve for this polynomial can only adopt one of the forms pictured on Fig. \ref{fig:4}. At any rate, $Q_2(X)=0$ has two purely real roots of opposite signs. The equation $Q_2(x)=0$ will then have two purely imaginary roots $\pm i\delta$, which shows in addition that the filamentation instability is here again purely growing with $\mathcal{R}\omega=0$. The growth rate $\delta$ is necessarily strictly positive unless $Q_2(0)=0$, that is $\theta=0$.


\end{document}